\documentclass{article}

\usepackage{PRIMEarxiv}

\usepackage[utf8]{inputenc} 
\usepackage[T1]{fontenc}    
\usepackage{hyperref}       
\usepackage{url}            
\usepackage{booktabs}       
\usepackage{amsfonts}       
\usepackage{nicefrac}       
\usepackage{microtype}      
\usepackage{lipsum}
\usepackage{arabtex}
\usepackage{amssymb}
\usepackage{utf8}
\usepackage{amsmath}

\usepackage[utf8]{inputenc} 
\usepackage[T1]{fontenc}    
\usepackage{hyperref}       
\usepackage{url}            
\usepackage{booktabs}       
\usepackage{amsfonts}       
\usepackage{nicefrac}       
\usepackage{microtype}      
\usepackage{lipsum}
\usepackage{graphicx}
\usepackage{multirow}
\usepackage{comment}

\usepackage{fancyhdr}       
\usepackage{graphicx}       
\setcode{utf8}
\begin{document}

\title{Arabic Text-To-Speech (TTS) Data Preparation
}

\author{
  Hala Al Masri \\
  Linguist Department \\
  Mawdoo3 Ltd \\
  Amman, Jordan\\
  \texttt{hala.masri@mawdoo3.com} \\
   \And
  Muhy Eddin Zater \\
  Electrical Engineering Department\\
  University of Colorado Boulder \\
  Boulder, Colorado\\
  \texttt{muhy.zater@colorado.edu} \\
}

\maketitle

    
People may be puzzled by the fact that voice over recordings data sets exist in addition to Text-to-Speech (TTS), Synthesis system advancements, albeit this is not the case. The goal of this study is to explain the relevance of TTS as well as the data preparation procedures. TTS relies heavily on recorded data since it can have a substantial influence on the outcomes of TTS modules. Furthermore, whether the domain is specialized or general, appropriate data should be developed to address all predicted language variants and domains. Different recording methodologies, taking into account quality and behavior, may also be advantageous in the development of the module. In light of the lack of Arabic language in present synthesizing systems, numerous variables that impact the flow of recorded utterances are being considered in order to manipulate an Arabic TTS module. In this study, two viewpoints will be discussed: linguistics and the creation of high-quality recordings for TTS. The purpose of this work is to offer light on how ground-truth utterances may influence the evolution of speech systems in terms of naturalness, intelligibility, and understanding. Well provide voice actor specs as well as data specs that will assist both voice actors and voice coaches in the studio as well as the annotators who will be evaluating the audios.



\section{Introduction and Related Work}

\subsection{Text-to-Speech Background and Overview}
Generating natural speech from text; despite decades of study, research and investigation, text-to-speech (TTS) synthesis remains a formidable challenge.\cite{taylor2009text}. Different techniques, implementations and algorithms were used for the task of generating speech from text. Starting with Concatenative Synthesis \cite{hunt1996unit} which is the process of connecting small units of waveforms together that were pre-recorded by humans\cite{black1997automatically} This approach was considered as the state-of-the-art for many years and was used in many fields and applications. Following the Concatenative Synthesis, Statistical parametric speech synthesis \cite{tokuda2000speech, tokuda2013speech, zen2009statistical, ze2013statistical} came to light, which generates smooth trajectories of speech features to be synthesized using a vocoder. This approach tackled various issues that the Concatenative Synthesis had. However, as opposed to human voice, the audio provided by the above devices frequently sounds muffled and unnatural.

The first attempt to generate human-like voice was WaveNet \cite{oord2016wavenet}, which is a generative model of time domain waveforms. WaveNet produces audio quality that mimics real human speech and is already used in some complete TTS systems. WaveNet makes use of certain linguistic characteristics, such as fundamental frequency and phoneme length, thus, it needs considerable domain knowledge to develop, as well as extensive text-analysis structures and a comprehensive lexicon.

Following WaveNet, researchers presented Tacotron \cite{wang2017tacotron} in order to be less dependent on domain experts. Tacotron is a sequence-to-sequence architecture that directly produces magnitude spectrograms from a sequence of characters, which simplifies the traditional speech synthesis pipeline by substituting the production of these linguistic and acoustic features with a single neural network trained from data alone, which in turn reduces the dependency on domain experts. To vocode the resulting magnitude spectrograms, Tacotron uses the Griffin-Lim algorithm for phase estimation, followed by an inverse short-time Fourier transform.

A hybridization of the best of the previous approaches uses an entire neural approach for speech synthesis; it utilizes sequence-to-sequence Tactotron-style model that generates mel spectrograms, followed by a modified WaveNet vocoder \cite{tamamori2017speaker}, and it was named Tactoron-2 \cite{shen2018natural}. Trained directly on normalized character sequences and corresponding speech waveforms, it learns to synthesize natural-sounding speech that is difficult to discern from actual human speech, producing cutting-edge text-to-speech results in a variety of languages.

\subsection{Arabic Text-to-Speech}

The Arabic language is the fifth most spoken and used language in the world and is used by different countries with over 500 Million people as a native language and hence it has gained an increased attention in the field of Natural Language Processing (NLP). On the other hand, research on Arabic language is challenging and poses other obstacles due to variety of reasons, including but not limited to; shortage of resources when it is compared to other languages like English or Spanish. Additionally, the Arabic language is considered as a morphologically rich and complex language, alongside the fact that it has many dialects that are significantly different from each other.

Researchers barely scratched the surface on Arabic TTS so far, with only few shy attempts to apply the aforementioned speech synthesis techniques on Arabic text \cite{zerrouki2019adapting, fahmy2020transfer, meriem2020phonetic}. However, Arabic TTS is  still far from what other languages have reached in terms of quality, naturalness and other traits. 

In the disclosure of invention, a complete End-to-End Arabic TTS that achieves state-of-the-art performance on Arabic TTS and generates a close to human voice is described and implemented. The intention of the usage of the disclosed TTS is for financial and banking fields and applications; hence more emphasis was applied on these domains as will be detailed in later sections.

The document is presented as following; Section shows usage examples of the presented system, where the following section illustrates a high-level look of the system and each of its components, whereas the six sections that follow describe in details each of the components used and implemented. Section presents the required data for implementation Last but not least, the final section describes the steps and measures considered in this disclosure to be more applicable to the financial domain.

\section{Text-to-Speech System Description}

The proposed text-to-speech system receives text as an input and generates speech as an output. A high level look on the overall system is illustrated in figure 1.

\begin{figure}[htbp]
\centering
\includegraphics[scale=0.6]{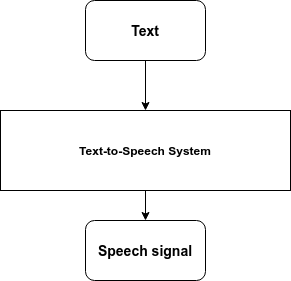}
\end{figure}

The shown TTS system is composed of multiple components that process and edit the input text followed by another component that generates audio given the processed text. Figure 2 below presents the contents of the TTS system.

\begin{figure}[htbp]
\centering
\includegraphics[scale=0.60]{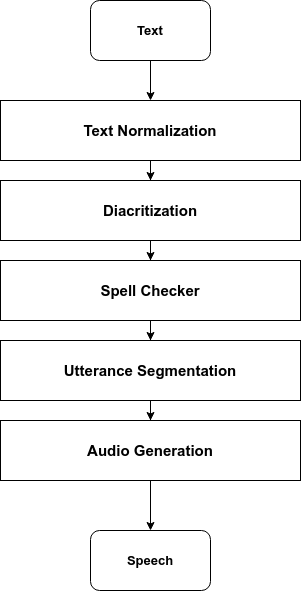}
\end{figure}

Each of the processes shown in the above figure plays an essential role in generating a correctly pronounced, human- like Arabic speech from text.

Briefly, the text normalization layer shown in figure 2 converts any numbers, dates non-Arabic words and abbreviations to Arabic written text. The layer that follows adds diacratics to the written text. This step is essential in generating a correctly pronounced Arabic speech. Furthermore, an Arabic spell checker is added to fix grammatical and spelling errors in the input text. Following is the utterance segmentation step, which chunks the input sentence into multiple meaningful sentences. Finally, the processed text is then generated into audio through the audio generation component. 
These layers and components will be discussed in details in later sections.

\section{Data Preparation}

The preparation of the acquired dataset, the selection of an attractive voice, the data validation procedure, and in-studio recording are all factors in the creation of Arabic TTS systems. Data preparation begins with data extraction based on the type of recorded data required, such as open domain data or close domain data. Data from the public domain (i.e. a public field content that is not protected by any copyright regulations and is a mixture of different topics but not limited to a specific genre). Closed domain data, on the other hand, is confined to a single field, such as finance, sport, law, names, and places.

Data preparation is often a difficult process. Regardless of the enormous work required, basic guidelines should be followed in order to obtain the greatest results. Annotation, on the other hand, is a critical component of the entire process. To address all conceivable spoken phonemic instances, data should be phonetically balanced. A balanced length each utterance is also important. Each phrase would include 12 utterances, with an average of 6 utterances over the whole data set. This is due to the fact that a normal individual could deliver a phrase made up of 12 utterances without pausing.

\subsection{Diacratization}

It is preferable to identify a ready-made diacritical data-set that is scarce or would be exceedingly expensive if available during the data production phase. Both the voice actor and the AI module would benefit from Arabic diacritics to minimize ambiguity. Because everyone understands how a single diacritic may change the entire meaning of a word or a phrase.

For example, if the term is diacritized, it might signify he studied, he taught someone, a noun that signifies a lesson, passive; was taught, and an ancient garment. The decision to add diacritics to any text is based on linguistic and contextual considerations. In the case of the TTS module, it would assist the machine in accurately reading the term without being confused about which one to use in a certain situation. Texts without diacritics can be difficult to read, especially for non-native Arabic speakers. A similar problem emerges with a module; it would be tough to contribute performance of various Natural Language Processing, NLP, for Arabic Applications.

However, diacritization may be done in a number of ways, depending on whether you want to diacritize partially or completely. Arabic manuscripts can be entirely diacritized, somewhat diacritized, or absent of diacritization. To overcome the uncertainty we described earlier, partial dicritization can be applied. Consider the following example.

\begin{figure}[h]
    \centering
    \includegraphics[width=4in, width=2in]{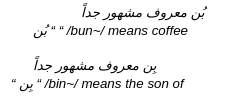}
    \caption{}
\end{figure}

As a result, a single diacritic has fundamentally changed the meaning of the sentence. There are eight diacritics in Arabic that can be used on a letter to denote its grammatical case. See the following table:

\begin{figure}[h]
    \centering
    \includegraphics[height=4in, width=4in]{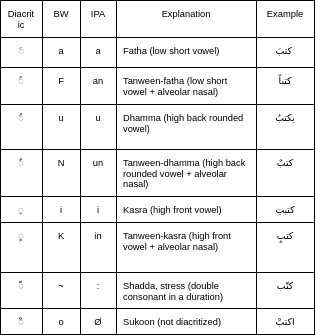}
    \caption{Table 1.  Arabic diacritics with their International Phonetic Alphabet IPA and Buckwalter BW equivalents}
\end{figure}

Few research publications use diacritics to show how unclear the texts are. Arabs are accustomed to writing sans diacritics, which makes Arabic text extremely confusing. Human readers arent the only ones who employ diacritics for clarification. It also applies to computerized equipment like morphological analyzers. To resolve ambiguity, certain morphological analyzers utilize partial diacritics, such as filtering solutions that are inconsistent with diacritics accessible in the input text.

\subsection{Recordings}

After diacritization, the next phase is in-studio recordings, which is the most important part of the whole process. In-studio recordings are essential for constructing TTS audios and maintaining consistency in producing a realistic, expressive, complete, understandable, and strong human-like voice. In reality, creating a human-like voice is difficult and still evolving.
TTS is made up of the following four modules:

\begin{itemize}
    \item Text preprocessing: a list of inputs and expected outputs should be produced at this step so that the real module outputs may be compared to the pre-expected ones. 
    \item Converting graphemes to phonemes: a module should translate utterances to a collection of phonemes, then calculate a percentage of correctness.
    \item Prosody computation: Phonemic strings and their accumulation should be completed.
    \item Voice waveforms: this stages module focuses on the creation of speech waveforms.

\end{itemize}

However, in-studio recordings are still required to train the module in its early phases and to preprocess the data processed. Selecting an Actor/Actress should be a competent professional voice over who is eager to work towards the same goals as you in order to produce a spectacular and successful production. [8] Gender: male or female, text type: narration, commercial, infomercials, chatbots, industry: banking, finance, healthcare, tone or style, and last but not least, special skills in terms of the ability to imitate a specific dialect or a native dialectical speaker should be listed when selecting the best talent.

Furthermore, the voice over talent should follow specific principles that, if followed, will result in the modules voice being created in a cohesive emotive human-like voice. Here are some data-specs guidelines to keep in mind when recording for TTS:

\subsubsection{Reading Speed}

Throughout the recording process, the tempo should remain consistent. This will maintain the high degree of consistency.

\subsubsection{General Flow}

\begin{itemize}
    \item Recording for no more than 2 hours each day is recommended. The key point of this point is that when working for more than 2-3 hours constantly, the voice over vocal acoustic system may become weary. S/he will eventually lose focus or get too stressed in his/her voice acoustic system. 
    \item In all recordings along similar lines, voices should be nice to the listener and not confrontational.
    \item Voices should be natural and not over-acted. The voice of the TTS might be made to sound like its from an animated film with a lot of effort and exaggeration.
    \item Returning to the reference sentences after 50 recording utterances to maintain a consistent speed throughout the procedure.
    \item It is strongly advised that you take a 10-minute break every hour. The vocal chords must be refreshed every hour.
    \item Technical cutbacks are just unacceptable. A better solution is repetition.

    \item To get back into the recording process, make a handful of reference phrases [2]. Reference Sentences will assist you in creating a sample that a voice over may rely on if the tone, pace, and other parameters of TTS recordings have been lost. Balance is essential phonologically.
    
\end{itemize}

\subsubsection{Pronunciation}

\begin{itemize}
    \item Be cautious while pronouncing velarized consonants such as:
    
    \begin{figure}[h]
    \centering
    \includegraphics[height=1in, width=1in]{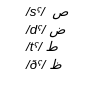}
\end{figure}
    
    This distinguishes them from their contemporaries:
    
    \begin{figure}[h]
    \centering
    \includegraphics[width=1in, width=1in]{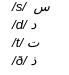}
\end{figure}

    Same is valid for the following consonants:
    
\begin{figure}[h]
    \centering
    \includegraphics[width=1in, width=1in]{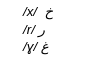}
\end{figure}    

    \item When /A/ hamzat wasl appears at the start of a phrase, pronounce it as if it were standard hamzat qate /gs/. Otherwise, it should be discarded.
    
    \item Short vowels should not be lengthened, and long vowels should not be shortened. In reading, give each vowel its own space.

    \item Without converting letters like /q/ to /g/ or /a/, the voice actor/ actress should endeavor to exhibit the required accent as much as feasible. According to MENA languages, Arabic is legally recognized by the government in 26 nations, with 18 countries having a majority of its population speaking it as their first language (Arab world where Arabic is the predominant language). There are around 30 significantly differing regional spoken variants (each nation has its own dialect) in addition to Modern Standard Arabic (MSA) in Arabic.
\end{itemize}

\subsubsection{Intonation}

\begin{itemize}
    \item Although earlier study has shown that pitch perceptions of tone language speakers would be better at low levels, very low levels would influence robotness difficulties, and high levels would impair the speed tempo.
    \item Intonation: The voice over actor/actress should have good intonation and be able to tell the difference between a full stop, a comma, and a question mark without overdoing it. Exaggerating would always result in narration, not a reliable TTS system.
    
    \item It is not advisable to change the voice based on the meaning of the statement. What important is that you pay attention to the punctuation marks.
\end{itemize}

\subsubsection{Pauses}

Pausing at punctuation marks is permitted; otherwise, no pauses are permitted.

Drop short vowels while halting; fatha //, dhamma //, kasra //, and tanween; alveolar nasal tanween-kasra / and tanween-dhamma /. Except for tanween-fatha on alif /a:/ + //, that is. When pausing, avoid saying tanween-fatha on alif to keep the text flowing, as doing so would change it into a long vowel, /a:/, as seen in the following example:

Drop diacritics (tashkeel) are mostly employed on ( ); subjective attached pronouns, such as /a:luk/,  even when it comes to feminine pronouns when it happens at the conclusion of a sentence.

\subsubsection{Background Noise}

When recording in an untreated environment, noise is unavoidable. Ultimately, the greatest method to eliminate noise is to prevent it from occurring in the first place. Hiss, rumble, crackle, hum, tongue pops, plosives, saliva, sibilance, and buzz are all examples of these sounds.

\section{Rating Speech Quality Criteria}

\subsection{Naturalness Test}

This test examines the naturalness of synthetic voices, or how near the voice is to that of a human. In this case, the evaluator must determine if the voice is natural, natural-sounding, or synthetic. A thorough cover is required to study such voices. This includes utterances, words, sentences, acronyms, abbreviations, numbers, and domains, among other things.

\subsection{Degraded Mean Opinion Score (MOS)}

This test compares the natural and synthetic voices to determine naturalness. Opinion testing, in which evaluators listen to a random selection of natural and synthetic audio recordings processed by different modules to ensure fair scoring, is a valuable tool during this phase. Voice human-likeness, domain coverage, and complexity are all factors to consider.

\begin{figure}[h]
    \centering
    \includegraphics[width=4in, width=2in]{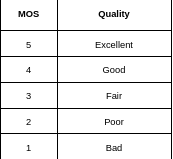}
\end{figure}

\begin{figure}[h]
    \centering
    \includegraphics[width=4in, width=2in]{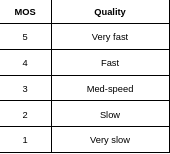}
\end{figure}

\subsection{Intelligibility Test}

This has an impact on speech quality and may be measured using the Mean Opinion Score (MOS) and Word Error Rate (WER) (WER). The first pertains to the precision with which any listener may comprehend the uttered utterance. The second test looks for phrases that are conceptually flawed yet syntactically correct. It means that the evaluator will write down what he or she heard, regardless of whether or not it was understood semantically. (Long-term memory people would be preferable.)

\subsection{Comprehensibility Test}

This test should be comprehended word for word. Evaluators will be asked to listen to a series of utterances and then answer a series of questions in order to determine if they can interpret words as tokens without understanding the meaning of the words. Answers would be scored as follows by evaluators as 0 for incorrect and 1 for correct.

Factors to be taken into consideration are shown herein:

\begin{itemize}
    \item Naturalness: Is the voice natural or artificial?
    \item Intelligibility (proper pronunciation): even if the words are logically erroneous, they are comprehensible.
    \item  Medium-tempo speech rate is preferred.
    \item Comprehensibility: receiving and comprehending the message.
    \item Efforts in listening: total relaxation is achievable; no efforts are necessary to listen appropriately.
    \item Friendliness: The tone of the voice should be pleasant.
    \item Robotism: It is not acceptable to be a robot.
    \item Consistency of the audios: how consistent are the audios?
    \item Overall impression: What is the evaluators overall impression of the files?
    \item Acceptance: Will the audios be accepted and usable for the TTS project?
\end{itemize}

\section{Conclusion}
Despite the fact that speech synthesis voice technologies are a rapidly evolving technology, there are still several constraints that complicate the approach of generating a human-like voice. Weve discovered a lack of consistency in Emotional TTS systems in terms of Arabic diacritics, data preparation and recording, and getting voice over talent ready to work. Aside from that, how can a suitable data collection be created to support TTS applications?
Weve gone through a set of rules that, if followed, would result in a rich emotive human-like voice from a voice over performer. On the other hand, the assumption that language and technological perspectives on a certain line of assessment metrics to be followed would lead to a more natural TTS would open the way.
The bulk of TTS Arabic synthesized systems are still weak and have similar flaws, as we can see. In conclusion, critical techniques have shed light on a newborn modules early toe steps, starting with unique data and finishing with created audios. The spectrum of the TTS system is broad, but linguistic enhancement should be prioritized in the foreseeable future.

\bibliographystyle{unsrt}  
\bibliography{references}

\end{document}